\documentclass[10pt,twocolumn,letterpaper]{article}

\usepackage[pagenumbers]{wacv} 

\usepackage{colortbl}
\usepackage{multirow}

\definecolor{wacvblue}{rgb}{0.21,0.49,0.74}
\usepackage[breaklinks,colorlinks,allcolors=wacvblue]{hyperref}

\def\wacvPaperID{1363} 
\def\confName{WACV}
\def\confYear{2027}

\title{SpatialTrust: A Benchmark for Environmental Risk Recognition in Secure Authentication}

\author{Junbin Lu, Hsiang-Wei Huang, Saesha Wadhwa, Yu Ting Hsu, Jenq-Neng Hwang\\
University of Washington, United States\\
{\tt\small \{junbinlu, hwhuang, saeshw, yhsu24, hwang\}@uw.edu}
}

\begin{document}
\maketitle
\begingroup
\renewcommand{\thefootnote}{}
\makeatletter
\def\Hy@footnote@currentHref{degree-note}
\makeatother
\footnotetext{Work done while the first author was at the University of Washington.}
\addtocounter{footnote}{-1}
\endgroup
\begin{abstract}
Visual environmental risk recognition plays an important role in secure authentication, where a user's surroundings may reveal sensitive information or introduce potential security risks. However, existing evaluations of multimodal large language models (MLLMs) rarely examine whether models can reliably recognize, localize, and explain such risks in spatially grounded authentication scenarios. We present SpatialTrust, a question-answering benchmark for evaluating environmental risk recognition in secure authentication. SpatialTrust assesses five complementary abilities: sensitive factor detection, direct factor identification, indirect factor identification, direct factor explanation, and indirect factor explanation. We evaluate both proprietary and open-source MLLMs and find that current models show limited performance, especially in understanding and explaining indirect risks, indicating that spatial risk awareness remains a challenging capability for MLLMs. In addition, we introduce SpatialTrustGuard, a structured QA-and-audit pipeline that improves Qwen3-VL-30B-A3B-Instruct from 36.78\% to 41.12\% overall. Our findings highlight the need for dedicated benchmarks and structured inference methods to improve the trustworthiness of MLLMs in secure authentication.
\end{abstract}
    
\section{Introduction}
\label{sec:intro}

Multimodal large language models (MLLMs) are increasingly used in secure authentication scenarios, including laptop-camera-based user verification, where a camera captures an image of the user together with part of the surrounding scene. In such settings, the captured image is usually intended for a narrow purpose, such as verifying that the correct user is present during login. However, because the camera also observes the environment around the user, the verification capture may contain visible background or non-primary-subject factors that could cause information leakage from the computer the user is logging into. For example, a nearby camera may face the login device, or another person may be positioned to observe the user's computer. Prior work on shoulder surfing and camera-based monitoring shows that physical surroundings can introduce privacy and security risks during authentication~\cite{aviv2017towards,balash2021examining}. Understanding whether MLLMs can recognize, explain, and localize these environmental risk factors is important for building trustworthy authentication systems.

Prior visual privacy datasets and tasks have focused on privacy attribute prediction and personalized privacy risk estimation~\cite{orekondy2017visual}, image-level privacy prediction~\cite{zhao2022privacyalert}, and private content in assistive images~\cite{gurari2019vizwizpriv}. More recent MLLM privacy and safety evaluations examine image-based safety attacks and harmful query scenarios~\cite{liu2024mmsafetybench} and individual-level privacy reasoning that links distributed multimodal evidence to identities~\cite{sun2025multipriv}. However, the risk considered in secure authentication is more specific: visible elements in the captured scene may disclose information from the login computer or create a risk of such disclosure during verification. This requires spatial and contextual reasoning about the relationship between the user, the camera, the login device, and the surrounding environment, as well as distinguishing direct disclosure from indirect risk. A reliable model should therefore identify risks based on visible evidence rather than unsupported assumptions about the user's identity, intent, occupation, or private context.

To address this gap, we introduce SpatialTrust, a question-answering benchmark for evaluating environmental risk recognition in secure authentication. SpatialTrust focuses on visible background or non-primary-subject factors that could cause information leakage from the computer the user is logging into during a verification capture. We define a direct sensitive factor as a visible factor that directly causes sensitive information to be disclosed in the current image. We define an indirect sensitive factor as a visible factor that does not directly disclose sensitive information in the current image, but creates a possibility or risk that sensitive information may be disclosed.

Each SpatialTrust item contains structured visual question-answering tasks that ask models to determine whether sensitive factors are present, distinguish direct from indirect sensitive factors, provide short natural-language explanations, and ground the relevant visual evidence with bounding boxes~\cite{zhu2016visual7w}. This design evaluates not only whether a model can select the correct answer, but also whether its answer is supported by concrete visible evidence in the image. By requiring both textual reasoning and spatial grounding, SpatialTrust tests whether MLLMs can produce evidence-based judgments about authentication-related environmental risks.

The SpatialTrust task is challenging for current MLLMs because models must focus on background or non-primary-subject factors rather than the verified user, distinguish direct disclosure from indirect risk, and avoid speculative inferences beyond visible evidence~\cite{li2023pope}. In addition, models must maintain consistency between selected options, textual explanations, and grounding boxes. These requirements make SpatialTrust a test of both environmental risk recognition and disciplined evidence-based response generation.

We systematically evaluate eight representative MLLMs~\cite{openai2026gpt54,comanici2025gemini25,vteam2025glm45v41v,zai2025glm46v,zhu2025internvl3,wang2024qwen2vl,bai2025qwen3vl} on SpatialTrust, covering both proprietary and open-source models. Our results reveal substantial limitations in current models when reasoning about environmental risks in authentication captures. In particular, models often struggle to distinguish direct and indirect sensitive factors, miss subtle risk factors in the background, or over-identify factors that are not sufficiently supported by visible evidence. Grounding also remains challenging, as models may produce answers whose visual evidence is incomplete, inconsistent, or poorly localized. These findings suggest that authentication-related environmental risk recognition remains an underdeveloped capability in current MLLMs.

In addition, we introduce SpatialTrustGuard, an inference-time QA-and-audit pipeline that decomposes answer selection, conservative auditing, and grounded response generation~\cite{zhou2023least,madaan2023selfrefine}. Built on Qwen3-VL-30B-A3B-Instruct~\cite{bai2025qwen3vl}, SpatialTrustGuard improves the overall score from 36.78\% to 41.12\%, suggesting that structured decomposition and answer auditing can improve model behavior, although environmental risk recognition in verification captures remains challenging.

Our contributions are threefold:
\begin{itemize}
    \item We introduce SpatialTrust, a benchmark for evaluating environmental risk recognition in secure authentication, with structured QA tasks covering sensitive factor detection, direct sensitive factor identification, indirect sensitive factor identification, direct sensitive factor explanation, and indirect sensitive factor explanation.
    \item We systematically evaluate eight representative MLLMs on SpatialTrust and reveal substantial limitations in current models, particularly in distinguishing direct and indirect sensitive factors and grounding relevant visual evidence.
    \item We propose SpatialTrustGuard, a staged inference-time QA pipeline with conservative self-verification, and show that structured decomposition and answer auditing can improve the overall performance of MLLMs on SpatialTrust.
\end{itemize}

\section{Related Work}
\label{sec:related}

\subsection{Privacy Risks in Authentication}

Visual privacy has been widely studied through tasks such as privacy attribute recognition, image privacy prediction, and privacy-sensitive region localization. Prior work has introduced visual privacy attributes and risk prediction~\cite{orekondy2017visual}, image-level privacy prediction datasets~\cite{zhao2022privacyalert}, and private visual information recognition in assistive settings~\cite{gurari2019vizwizpriv}. Authentication research further shows that privacy and security risks can arise from the user's physical environment: shoulder-surfing can leak authentication information through observation of devices or input behavior~\cite{aviv2017towards}, while camera-based monitoring may capture sensitive information about the user and surrounding scene~\cite{balash2021examining,mukherjee2024balancing}.

These works provide important foundations, but they usually study general private content, specific attack procedures, or user perceptions of monitoring. In contrast, SpatialTrust focuses on laptop-camera-based login verification, where background or non-primary-subject factors such as observers or cameras may leak information from the login computer. SpatialTrust therefore requires models to recognize visible risk evidence and judge whether it creates a direct or indirect information-leakage risk in the verification capture.

\subsection{Privacy and Safety Evaluation in VLMs}

Recent vision-language models (VLMs) and multimodal large language models (MLLMs) have achieved strong performance on general visual reasoning tasks, but their privacy and safety behavior remains a concern. Existing benchmarks evaluate MLLM safety under image-based manipulations and harmful query scenarios~\cite{liu2024mmsafetybench} and individual-level privacy reasoning through identity linkage across multimodal evidence~\cite{sun2025multipriv}. However, these evaluations mainly focus on harmful content or linking identifiers to personal attributes. SpatialTrust complements them by evaluating whether VLMs can identify authentication-related environmental risks, distinguish direct from indirect sensitive factors, and support their answers with visible evidence.

\section{SpatialTrust Benchmark}
\label{sec:benchmark}

We introduce SpatialTrust, a benchmark for evaluating whether multimodal large language models can recognize spatially grounded information-leakage risks in user verification scenarios. The scenario is a laptop-camera-based verification setting, where the camera captures an image of the user and the surrounding scene while the user is logging into a computer. SpatialTrust focuses on visible background or non-primary-subject factors that could cause information leakage from the computer the user is logging into during this verification capture.

This setting is different from general visual privacy recognition. The goal is not to determine whether an image contains private content in a broad sense, nor to treat the primary user as a sensitive factor. Instead, SpatialTrust asks whether visible factors outside the primary subject may expose, capture, reflect, or enable access to information from the login computer. Therefore, the task requires models to reason about the spatial relationship among the user, the laptop camera, the login computer, and the surrounding environment.

\subsection{Task Definition}
\label{sec:task_definition}

\begin{figure*}[!t]
\centering
\includegraphics[width=\textwidth]{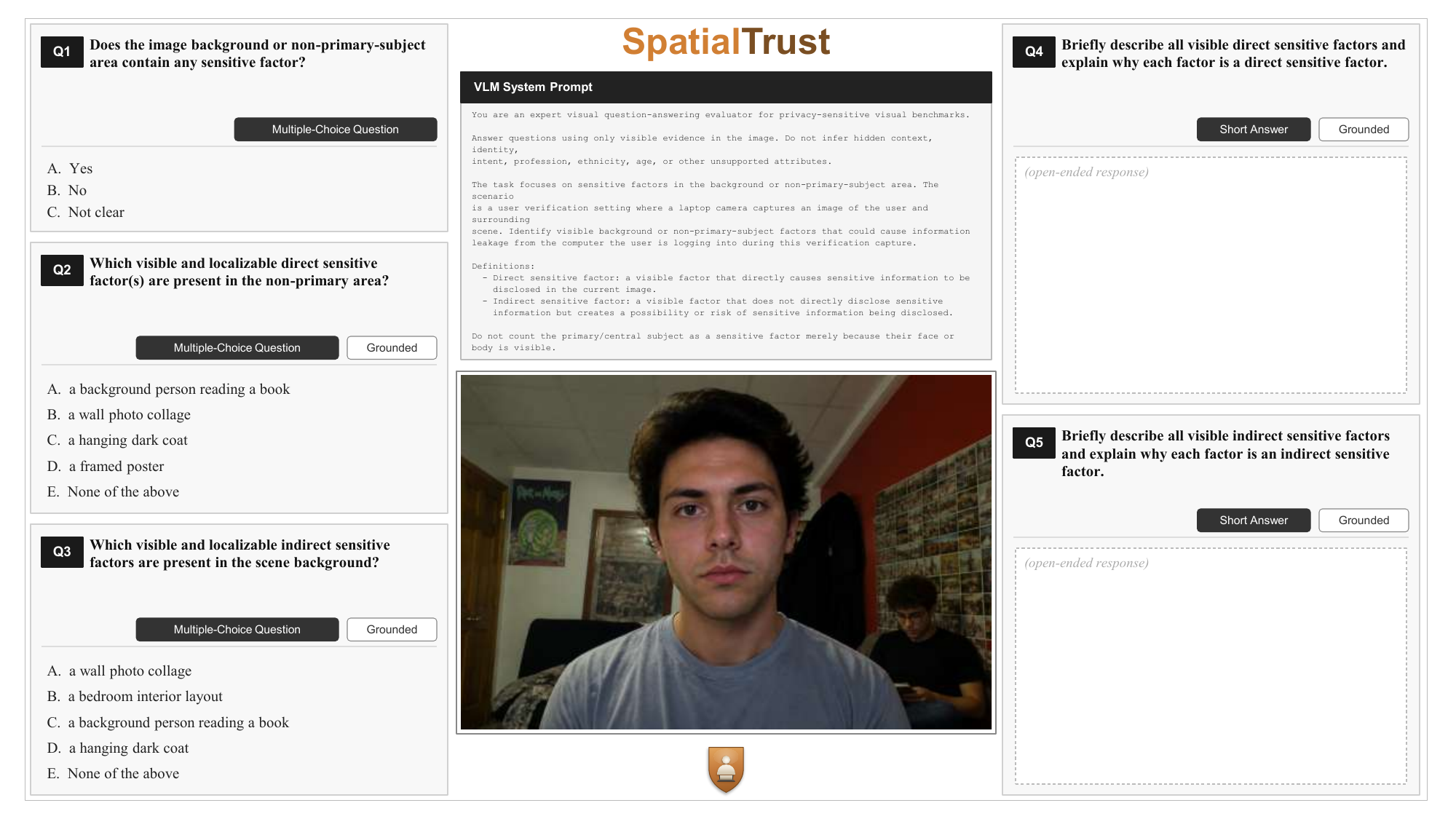}
\caption{\textbf{Example SpatialTrust item.} Each image is paired with a visible-evidence-only system prompt and five structured questions.}
\label{fig:spatialtrust_example}
\end{figure*}

Given an image captured during user verification, the model must identify visible background or non-primary-subject factors that could cause information leakage from the computer the user is logging into. Each benchmark instance contains five questions, covering sensitive factor detection, direct sensitive factor identification, indirect sensitive factor identification, direct factor explanation, and indirect factor explanation, as illustrated in \cref{fig:spatialtrust_example}.

All answers must be based only on visible evidence in the image. A valid prediction should not infer hidden intent, user identity, occupation, unseen activity, or private context beyond the image. When a sensitive factor is visible and localizable, the model is required to provide a bounding box for the supporting evidence. This design evaluates not only whether a model gives the correct answer, but also whether its judgment is grounded in the correct background or non-primary-subject region.

\subsection{Direct and Indirect Sensitive Factors}
\label{sec:direct_indirect}

SpatialTrust separates sensitive factors into direct and indirect categories. A direct sensitive factor is a visible factor that directly causes sensitive information to be disclosed in the current image. In the user verification setting, this means that the factor already creates an immediate disclosure path in the captured scene. Examples include a non-primary person looking toward the camera or login area, a phone or camera aimed at the scene, or a surveillance camera facing the user.

An indirect sensitive factor is a visible factor that does not directly disclose sensitive information in the current image, but creates a possibility or risk of sensitive information being disclosed. Examples include an open doorway that could allow another person to observe the login computer or screen area, a window or layout that increases information exposure risk, or a nearby person who is not currently observing the screen but could potentially do so.

This distinction is central to SpatialTrust. Direct factors correspond to visible evidence of immediate disclosure, while indirect factors correspond to visible conditions that may enable disclosure. At the same time, the benchmark requires conservative reasoning: a factor should be labeled sensitive only when the image provides concrete visual evidence supporting that judgment.

\subsection{Data Collection and Annotation}
\label{sec:data_collection}

\begin{figure*}[!t]
\centering
\includegraphics[width=\textwidth]{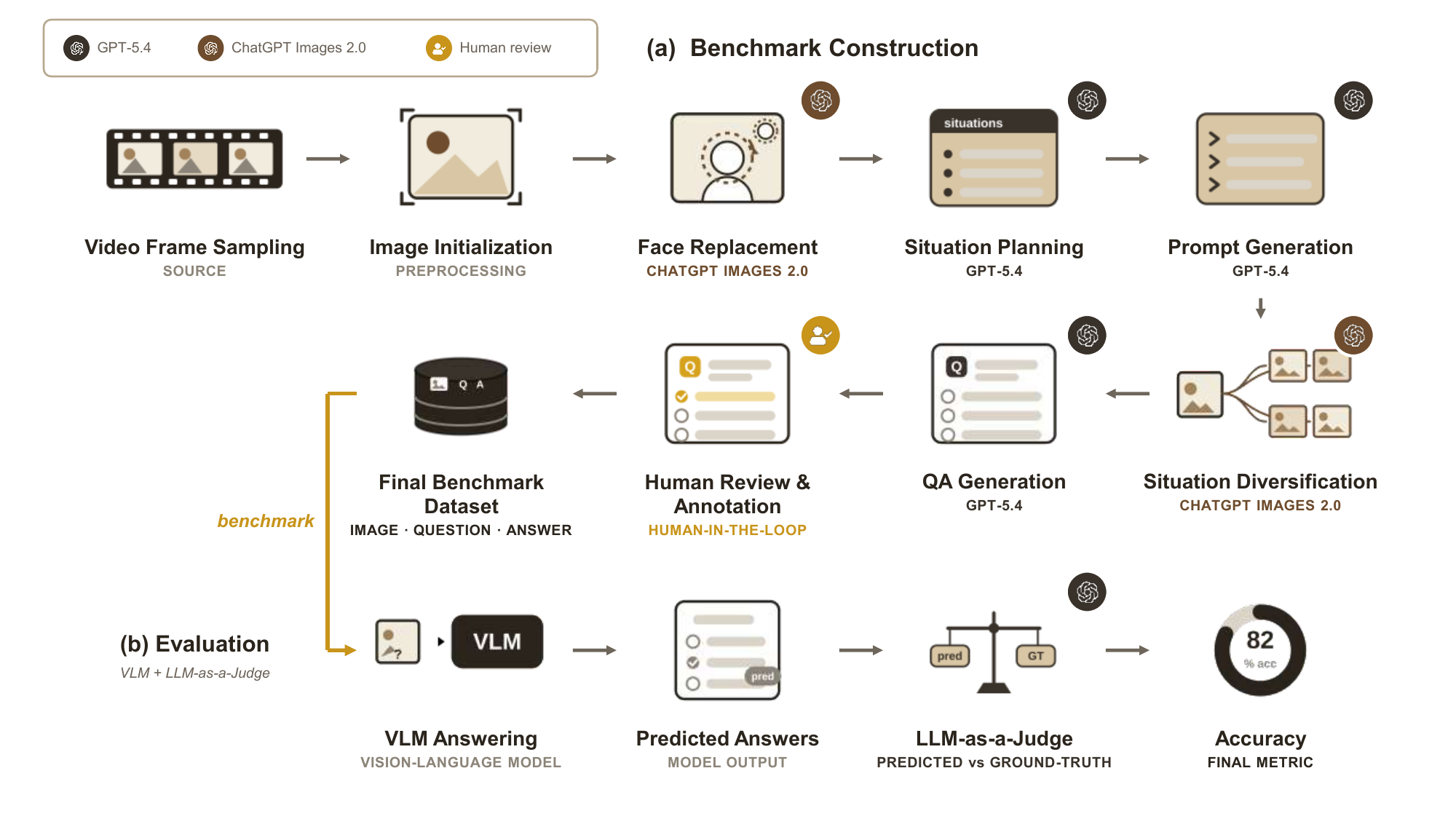}
\caption{\textbf{Overview of the SpatialTrust benchmark construction and evaluation pipeline.} The benchmark is built through video-frame sampling, image preprocessing, identity-level and situation-level image editing, QA generation, and human review; evaluation compares model predictions against ground truth using answer accuracy and grounded judging.}
\label{fig:benchmark_pipeline}
\end{figure*}

\cref{fig:benchmark_pipeline} shows the overall benchmark construction and evaluation pipeline. SpatialTrust images are constructed from frames sampled from a source video~\cite{youtube_spatialtrust_source}. We first apply initial preprocessing to the selected frames, including image-size normalization and other standardization operations, so that the images follow a consistent authentication-style format. Each processed frame contains a central user and a surrounding scene, with the benchmark focusing on background and non-primary-subject regions.

\begin{figure}[!t]
\centering
\includegraphics[width=\linewidth]{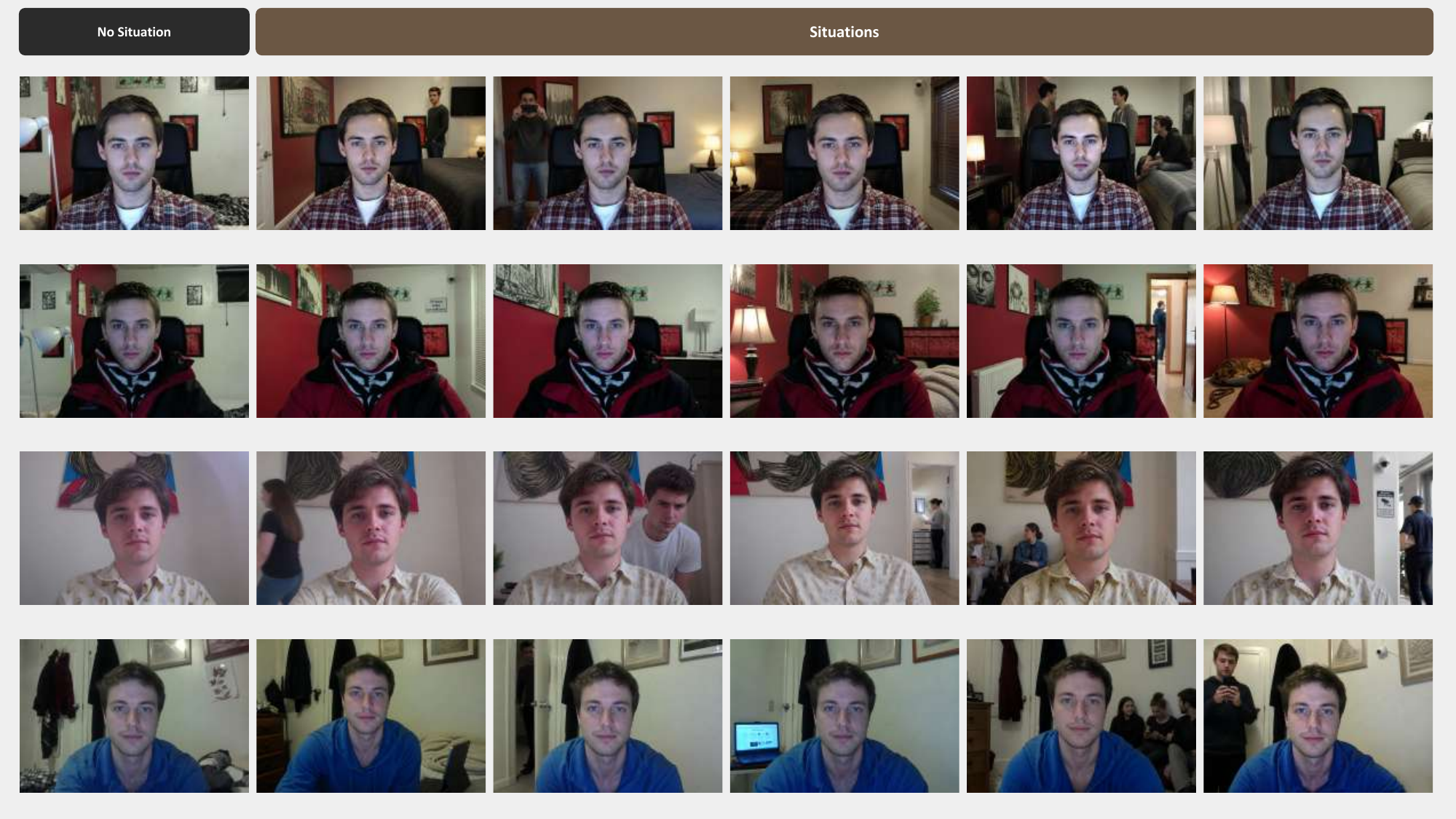}
\caption{\textbf{Examples of situation-level diversification in SpatialTrust.} The editing stage produces both images without an added sensitive situation and images with varied authentication-related sensitive-factor configurations.}
\label{fig:situation_diversification}
\end{figure}

We then use ChatGPT Images 2.0~\cite{openai2026chatgptimages} to edit the preprocessed images. The central person in each image is replaced through face swapping, and if other faces appear in non-central regions, those faces are also replaced. After this identity-level editing step, GPT-5.4~\cite{openai2026gpt54} analyzes each identity-edited image, designs situation-level editing plans, and generates the corresponding image-editing prompts. We then call ChatGPT Images 2.0~\cite{openai2026chatgptimages} again with these prompts to diversify the images at the situation level, creating varied authentication scenarios and sensitive-factor configurations, as illustrated in \cref{fig:situation_diversification}. Finally, GPT-5.4~\cite{openai2026gpt54} is used to generate the fixed questions, answer choices, and initial answers for each image. Human annotators then manually review the generated content, revise the answer choices, and label the correct answers to ensure the quality and reliability of the benchmark.

\subsection{Dataset Statistics}
\label{sec:dataset_statistics}

\cref{tab:dataset_statistics} summarizes the main statistics of SpatialTrust. The benchmark contains 577 images and 2,885 QA items, with five questions for each image. All images are normalized to a resolution of $1440 \times 960$. The QA set contains 577 answer-only questions and 2,308 grounded questions. By question type, the benchmark includes 1,612 single-choice questions, 119 multiple-choice questions, and 1,154 short-answer questions.

At the image level, 391 images contain at least one sensitive factor and 186 images contain no sensitive factor. For localizable sensitive-factor annotations, 220 images contain direct sensitive factors and 280 images contain indirect sensitive factors. These two categories are not mutually exclusive: 109 images contain both direct and indirect sensitive factors, 111 contain only direct factors, 171 contain only indirect factors, and 186 contain neither type.

SpatialTrust also provides grounding annotations for evaluating whether model predictions are supported by the correct visual evidence. In total, 1,000 QA items require evidence boxes, and the dataset contains 1,262 annotated evidence boxes. All boxes use absolute \texttt{xyxy} coordinates and fall within the image boundaries. These annotations allow SpatialTrust to evaluate both the correctness of model answers and the spatial grounding of the supporting evidence.

\begin{table}[t]
\centering
\caption{\textbf{Dataset statistics of SpatialTrust.}}
\label{tab:dataset_statistics}
\resizebox{\columnwidth}{!}{%
\begin{tabular}{l|c}
\hline
Statistic & Count \\
\hline
Images & 577 \\
Image resolution & $1440 \times 960$ \\
QA items & 2,885 \\
QA items per image & 5 \\
Answer-only QA items & 577 \\
Grounded QA items & 2,308 \\
Single-choice questions & 1,612 \\
Multiple-choice questions & 119 \\
Short-answer questions & 1,154 \\
\hline
Images with any sensitive factor & 391 \\
Images without sensitive factors & 186 \\
Images with direct sensitive factors & 220 \\
Images with indirect sensitive factors & 280 \\
Images with both direct and indirect sensitive factors & 109 \\
Images with only direct factors & 111 \\
Images with only indirect factors & 171 \\
Images with neither factor type & 186 \\
\hline
QA items requiring evidence boxes & 1,000 \\
Annotated evidence boxes & 1,262 \\
\hline
\end{tabular}
}
\end{table}

\subsection{Evaluation Protocol}
\label{sec:evaluation_protocol}

We evaluate all models on a fixed benchmark of 577 images, where each image is associated with five questions covering both answer-only recognition and grounded reasoning. The questions are evaluated at the image-question level. q1 measures whether the model can correctly determine whether the non-primary-subject region contains any sensitive factor. q2 and q3 evaluate grounded recognition of direct and indirect sensitive factors, respectively. q4 and q5 further evaluate whether the model can describe the corresponding direct and indirect sensitive factors and explain why they are sensitive.

For q1, we use exact-match accuracy over the normalized multiple-choice answer. For q2 and q3, we use option-level scoring with grounding verification. The predicted option set is first compared with the ground-truth option set. If the prediction contains any extra incorrect option, the score for that question is set to zero. Otherwise, each correctly predicted ground-truth option receives credit when it is sufficiently grounded. For options without a corresponding ground-truth evidence box, such as \texttt{None of the above}, selecting the option is sufficient because no IoU can be assessed.

The grounding quality is measured using Intersection over Union (IoU). Given a predicted box $B_p$ and a ground-truth box $B_g$, IoU is defined as:
\begin{equation}
\begin{aligned}
\mathrm{IoU}(B_p, B_g)
&=
\frac{|B_p \cap B_g|}
{|B_p \cup B_g|} \\
&=
\frac{|B_p \cap B_g|}
{|B_p| + |B_g| - |B_p \cap B_g|}.
\end{aligned}
\end{equation}

We use an IoU threshold of $\tau=0.5$ in all experiments. Let $G$ and $P$ denote the ground-truth and predicted option sets, respectively. For each option $o$, let $\mathcal{B}_g(o)$ and $\mathcal{B}_p(o)$ be the sets of ground-truth and predicted evidence boxes associated with that option. We define
\begin{equation}
M(o)=
\max_{B_p \in \mathcal{B}_p(o),\, B_g \in \mathcal{B}_g(o)}
\mathrm{IoU}(B_p,B_g),
\end{equation}
with $M(o)=0$ when either box set is empty. For q2 and q3, the question score is
\begin{equation}
s_{2/3} =
\begin{cases}
0, & P \setminus G \neq \emptyset, \\
\frac{1}{|G|}\sum_{o \in G} c(o), & \text{otherwise},
\end{cases}
\end{equation}
where
\begin{equation}
c(o)=
\begin{cases}
1, & o \in P \text{ and } \mathcal{B}_g(o)=\emptyset, \\
1, & o \in P \text{ and } \mathcal{B}_g(o)\neq\emptyset \text{ and } M(o)>\tau, \\
0, & \text{otherwise}.
\end{cases}
\end{equation}
Thus, q2 and q3 use the maximum IoU between predicted and ground-truth boxes under the same option label rather than requiring a fixed one-to-one box correspondence.

As shown in the evaluation stage of \cref{fig:benchmark_pipeline}, q4 and q5 require free-form textual answers, so we combine semantic evaluation with grounding evaluation. A structured LLM judge~\cite{zheng2023judging} compares the model response with the ground-truth answer and determines whether each ground-truth sensitive factor is semantically identified and whether the explanation correctly states why the factor is sensitive. The judge also flags unsupported predictions, where the model claims sensitive factors that are not present in the ground truth; such cases receive zero score for the corresponding question.

For q4 and q5, let $u$ indicate whether the judge flags an unsupported prediction. Let $N$ be the number of ground-truth evidence annotations for the question. When $N>0$, the judge matches each ground-truth factor $i$ to at most one predicted evidence box $B_{p(i)}$. Let $a_i$ indicate semantic correctness and $e_i$ indicate explanation correctness for factor $i$. We credit a factor only when it is semantically correct, its explanation is correct, and its matched evidence box is sufficiently grounded:
\begin{equation}
r_i =
\mathbf{1}\!\left[
a_i=1 \land e_i=1 \land
\mathrm{IoU}(B_{p(i)},B_{g,i})>\tau
\right].
\end{equation}
If the matched predicted evidence box is missing or invalid, $r_i=0$. The q4 and q5 score is
\begin{equation}
s_{4/5} =
\begin{cases}
0, & u=1, \\
r_0, & u=0 \text{ and } N=0, \\
\frac{1}{N}\sum_{i=1}^{N} r_i, & u=0 \text{ and } N>0,
\end{cases}
\end{equation}
where $r_0=1$ only if the model clearly states that no such factor exists and provides no spurious evidence; otherwise $r_0=0$.

\section{SpatialTrustGuard}
\label{sec:spatialtrustguard}

We propose SpatialTrustGuard, an inference-time strategy that improves model consistency on SpatialTrust without fine-tuning or additional training data. The method targets a common failure mode in grounded privacy-risk reasoning: models may introduce unsupported sensitive factors and then generate fluent explanations and plausible-looking grounding regions for them. SpatialTrustGuard addresses this issue by decomposing the task into staged answer selection, conservative verification, and final grounded generation, as detailed below.

\subsection{Method Overview}
\label{sec:guard_overview}

\begin{figure}[!t]
\centering
\includegraphics[width=\linewidth]{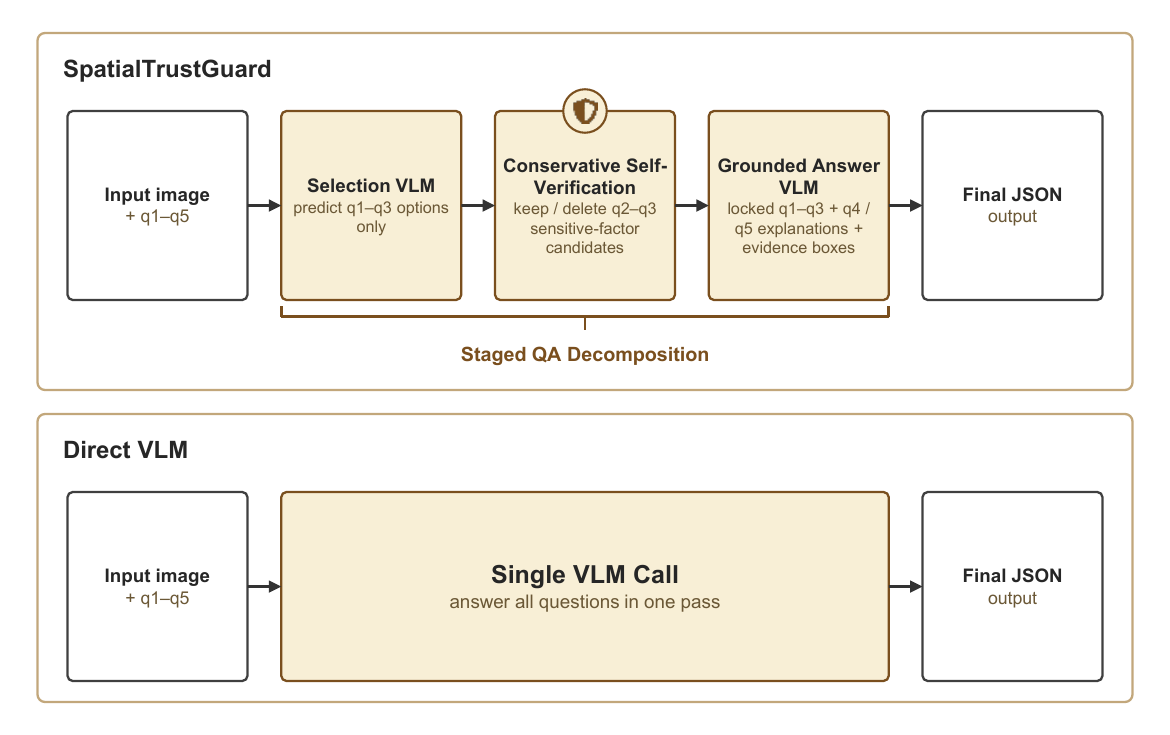}
\caption{\textbf{Overview of SpatialTrustGuard.} Compared with direct one-pass VLM answering, SpatialTrustGuard decomposes inference into candidate selection, conservative self-verification, and grounded answer generation with locked intermediate choices.}
\label{fig:spatialtrustguard_pipeline}
\end{figure}

Given an image $I$ and the five SpatialTrust questions $Q=\{q_1,\ldots,q_5\}$, SpatialTrustGuard produces a structured answer set $A=\{a_1,\ldots,a_5\}$. As shown in \cref{fig:spatialtrustguard_pipeline}, the pipeline contains three stages.

First, the model performs candidate factor selection for $q_1$, $q_2$, and $q_3$. This stage asks only for categorical judgments: whether any sensitive factor is present, which direct sensitive factors are visible, and which indirect sensitive factors are visible. By postponing explanation and localization, this stage isolates the model's basic recognition decision from the more open-ended generation tasks that follow.

Second, SpatialTrustGuard applies conservative self-verification~\cite{madaan2023selfrefine} to the selected $q_2$ and $q_3$ options. The verifier re-examines each proposed sensitive factor under an evidence-sufficiency criterion: the factor should be retained only if the complete described condition is clearly visible in the image. The verifier is restricted to removing unsupported candidates and is not allowed to introduce new factors. This makes the verification step conservative and reduces the chance that later explanations will be built around speculative claims.

Third, the model generates the final grounded answers for all five questions using the verified $q_1$--$q_3$ answers as fixed semantic commitments. The direct-factor explanation in $q_4$ must be derived from the verified direct factors in $q_2$, and the indirect-factor explanation in $q_5$ must be derived from the verified indirect factors in $q_3$. When no visually supported factor is retained for a category, the final response states the absence of such evidence rather than inventing a grounding target. The generated answer is then checked for structural and semantic consistency, including preservation of the verified choices and alignment between identification labels, explanations, and grounding regions.

\subsection{Staged QA Decomposition}
\label{sec:staged_decomposition}

The first stage decomposes SpatialTrust into a lightweight choice-answering task. The model receives the image together with the question text and answer options for $q_1$, $q_2$, and $q_3$, while ground-truth annotations, explanations, and grounding regions are withheld. The output of this stage is a compact set of selected option labels, with a single choice for the binary presence question and a set of choices for each factor-identification question.

This decomposition reduces the burden of solving several different tasks at once~\cite{zhou2023least}. In a one-pass prompting setting, the model must simultaneously detect sensitive factors, distinguish direct from indirect factors, write explanations, and localize evidence. These requirements can interfere with each other. For example, the model may select one factor in $q_2$, describe a different factor in $q_4$, or produce a bounding box that corresponds to neither. By separating answer selection from explanation and grounding, SpatialTrustGuard first establishes a compact answer state that later stages must follow.

The selected answers are standardized into a compact choice state before grounded generation. SpatialTrustGuard enforces a single valid choice for $q_1$ and valid sets of choices for $q_2$ and $q_3$, with mutually exclusive options such as ``None of the above'' handled as exclusive selections. After verification, the final stage receives the full benchmark item together with these fixed answers: grounding is generated only for verified $q_2$ and $q_3$ factor choices, and the explanations in $q_4$ and $q_5$ must remain tied to the same direct and indirect factors, respectively. If no factor is verified for a category, the response records the absence of visible supporting evidence and assigns no grounding region.

\subsection{Conservative Self-Verification}
\label{sec:conservative_verification}

The conservative self-verification stage audits the sensitive-factor options selected in $q_2$ and $q_3$. For each selected non-``None of the above'' option, the verifier evaluates whether the complete option description is visually supported by the image. An option is kept only when the relevant factor is clearly visible and does not require inference beyond the available visual evidence.

The verifier is intentionally restricted to deletion-only editing. It cannot add a missed sensitive factor, replace an incorrect factor with another option, or alter $q_1$. This restriction mainly targets false positives, which are especially harmful in SpatialTrust because an unsupported factor can later be expanded into a fluent explanation and paired with an arbitrary grounding region. By allowing only removal, the verifier reduces unsupported claims while preserving the original selection structure.

If all proposed factors for $q_2$ or $q_3$ are removed, the corresponding category is treated as having no visually supported sensitive factor. The verified choices then serve as fixed constraints for final generation. This prevents the final stage from introducing new direct or indirect factors that were not selected and verified.

\begin{table*}[!t]
\centering
\caption{\textbf{Model performance on SpatialTrust} with an IoU threshold of 0.5 for grounded evidence.}
\label{tab:vlm_performance}
\setlength{\tabcolsep}{7pt}
\renewcommand{\arraystretch}{1.15}
\setlength{\aboverulesep}{0pt}
\setlength{\belowrulesep}{0pt}
\resizebox{\textwidth}{!}{
\begin{tabular}{@{}lcccccc@{}}
\toprule
\multirow{2}{*}[-0.5ex]{\textbf{Model}} &
\multirow{2}{*}[-0.5ex]{\shortstack{\textbf{Sensitive Factor}\\\textbf{Detection}}} &
\multicolumn{2}{c}{\rule[-1.3ex]{0pt}{4ex}\textbf{Factor Identification}} &
\multicolumn{2}{c}{\rule[-1.3ex]{0pt}{4ex}\textbf{Factor Explanation}} &
\multirow{2}{*}[-0.5ex]{\textbf{Overall}} \\
\cmidrule(lr){3-4}\cmidrule(lr){5-6}
& & \rule[-1.3ex]{0pt}{4ex}Direct & Indirect & Direct & Indirect & \\
\midrule
\rowcolor[gray]{0.92}\multicolumn{7}{@{}l}{\textbf{\textit{Proprietary Models}}} \\
GPT-5.4 & 68.80 & 26.08 & 5.83 & 24.44 & 4.53 & 25.94 \\
Gemini-2.5-flash & 66.38 & 40.81 & 12.36 & \textbf{38.91} & 10.60 & 33.81 \\
\midrule
\rowcolor[gray]{0.92}\multicolumn{7}{@{}l}{\textbf{\textit{Open-Source Models}}} \\
GLM-4.1V-9B-Thinking & 67.07 & 14.21 & 8.64 & 11.96 & 6.21 & 21.62 \\
GLM-4.6V-Flash & \textbf{69.32} & 26.78 & 13.75 & 15.94 & 6.47 & 26.45 \\
InternVL3-38B-Instruct & 67.76 & 9.10 & 4.74 & 6.41 & 3.52 & 18.31 \\
Qwen2-VL-7B-Instruct & 66.38 & 6.76 & 3.90 & 1.56 & 0.87 & 15.89 \\
Qwen3-VL-8B-Instruct & 68.80 & 28.68 & 17.85 & 18.28 & 10.92 & 28.91 \\
Qwen3-VL-30B-A3B-Instruct & 67.42 & 39.17 & 28.65 & 25.74 & 22.91 & 36.78 \\
\midrule
\rowcolor[gray]{0.92}\multicolumn{7}{@{}l}{\textbf{\textit{Our Method}}} \\
\textbf{SpatialTrustGuard} & 65.51 & \textbf{41.07} & \textbf{34.20} & 35.44 & \textbf{29.38} & \textbf{41.12} \\
\bottomrule
\end{tabular}
}
\end{table*}

\begin{table*}[!t]
\centering
\caption{\textbf{Ablation study of SpatialTrustGuard} with an IoU threshold of 0.5 for grounded evidence.}
\label{tab:ablation}
\setlength{\tabcolsep}{7pt}
\renewcommand{\arraystretch}{1.15}
\setlength{\aboverulesep}{0pt}
\setlength{\belowrulesep}{0pt}
\resizebox{\textwidth}{!}{
\begin{tabular}{@{}lcccccc@{}}
\toprule
\multirow{2}{*}[-0.5ex]{\textbf{Variant}} &
\multirow{2}{*}[-0.5ex]{\shortstack{\textbf{Sensitive Factor}\\\textbf{Detection}}} &
\multicolumn{2}{c}{\rule[-1.3ex]{0pt}{4ex}\textbf{Factor Identification}} &
\multicolumn{2}{c}{\rule[-1.3ex]{0pt}{4ex}\textbf{Factor Explanation}} &
\multirow{2}{*}[-0.5ex]{\textbf{Overall}} \\
\cmidrule(lr){3-4}\cmidrule(lr){5-6}
& & \rule[-1.3ex]{0pt}{4ex}Direct & Indirect & Direct & Indirect & \\
\midrule
Qwen3-VL-30B-A3B-Instruct & \textbf{67.42} & 39.17 & 28.65 & 25.74 & 22.91 & 36.78 \\
+ Staged QA Decomposition & 66.38 & 38.13 & 31.57 & 31.72 & 27.01 & 38.96 \\
+ Conservative Self-Verification & 65.51 & \textbf{41.07} & \textbf{34.20} & \textbf{35.44} & \textbf{29.38} & \textbf{41.12} \\
\bottomrule
\end{tabular}
}
\end{table*}

\section{Experiments}
\label{sec:experiments}

\subsection{Experimental Setup}
\label{sec:experimental_setup}

We evaluate multimodal large language models on the full SpatialTrust benchmark using the image and five structured questions defined in \cref{sec:task_definition}. For localization questions, evidence boxes are scored with an IoU threshold of 0.5.

We report five task-level scores: sensitive factor detection, direct factor identification, indirect factor identification, direct factor explanation, and indirect factor explanation. The overall score is the average performance over all five question types. All scores are reported as percentages. We compare two proprietary models, GPT-5.4~\cite{openai2026gpt54} and Gemini-2.5-flash~\cite{comanici2025gemini25}, with six open-source models: GLM-4.1V-9B-Thinking~\cite{vteam2025glm45v41v}, GLM-4.6V-Flash~\cite{zai2025glm46v}, InternVL3-38B-Instruct~\cite{zhu2025internvl3}, Qwen2-VL-7B-Instruct~\cite{wang2024qwen2vl}, Qwen3-VL-8B-Instruct and Qwen3-VL-30B-A3B-Instruct~\cite{bai2025qwen3vl}. SpatialTrustGuard is built on Qwen3-VL-30B-A3B-Instruct and uses the staged inference pipeline described in \cref{sec:spatialtrustguard}.

\subsection{Main Results on SpatialTrust}
\label{sec:main_results}

\cref{tab:vlm_performance} shows the performance of representative VLMs and SpatialTrustGuard on SpatialTrust. Current VLMs achieve moderate sensitive factor detection, with most models scoring around 66--69\%, but their performance drops substantially on factor identification, explanation, and grounding. This indicates that detecting possible sensitive factors is easier than localizing the relevant evidence and explaining its role in authentication-related information leakage.

Among proprietary models, Gemini-2.5-flash outperforms GPT-5.4 overall, mainly due to stronger factor identification and explanation. Among open-source models, Qwen3-VL-30B-A3B-Instruct performs best, reaching 36.78\% overall and showing relatively strong results on indirect factor identification and explanation. Qwen3-VL-8B-Instruct and GLM-4.6V-Flash obtain the next best open-source results, but still trail the larger Qwen3 model on the more difficult indirect and explanation tasks. In contrast, InternVL3-38B-Instruct and Qwen2-VL-7B-Instruct achieve comparable sensitive factor detection scores but much lower grounded explanation scores, suggesting that general detection ability does not directly translate into reliable spatial reasoning. Overall, the open-source results show that SpatialTrust is less limited by recognizing whether a sensitive factor may exist and more limited by identifying the correct factor type and grounding the explanation.

SpatialTrustGuard achieves the best overall score, improving Qwen3-VL-30B-A3B-Instruct from 36.78\% to 41.12\%. The gains mainly come from structured reasoning and grounding: indirect factor identification improves from 28.65\% to 34.20\%, direct factor explanation from 25.74\% to 35.44\%, and indirect factor explanation from 22.91\% to 29.38\%. Sensitive factor detection drops from 67.42\% to 65.51\%, reflecting the conservative deletion-only verification step, but the overall score improves because SpatialTrust emphasizes correct identification, explanation, and grounding of direct and indirect evidence.

\subsection{Ablation Study}
\label{sec:ablation_study}

\cref{tab:ablation} evaluates the contributions of Staged QA Decomposition and Conservative Self-Verification. Starting from the Qwen3-VL-30B-A3B-Instruct one-pass baseline, we first add Staged QA Decomposition, which separates candidate selection from final grounded answer generation while omitting self-verification. This raises the overall score from 36.78\% to 38.96\%, showing that decomposing SpatialTrust into intermediate answer states helps produce more consistent final responses.

Adding Conservative Self-Verification on top of this decomposition further raises the overall score to 41.12\%. The full SpatialTrustGuard pipeline outperforms the decomposition-only variant on direct and indirect factor identification and both explanation tasks. The largest gains occur in explanation and grounding, where locked and verified choices constrain the final response to evidence tied to the selected sensitive factors. These results show that both decomposition and conservative answer auditing are useful, with verification providing further protection against unsupported factor claims.

\section{Conclusion}
\label{sec:conclusion}

We introduced SpatialTrust, a benchmark for evaluating whether multimodal large language models can recognize, distinguish, explain, and ground environmental information-leakage risks in secure authentication captures. SpatialTrust focuses on visible background and non-primary-subject factors that may disclose, reflect, record, or enable access to information from the login computer, and evaluates whether model judgments are supported by concrete visual evidence. Our evaluation of eight representative MLLMs shows that current models remain limited, especially in distinguishing direct and indirect sensitive factors and grounding their explanations in relevant evidence. We further proposed SpatialTrustGuard, a staged inference-time pipeline with conservative self-verification, which improves the overall score from 36.78\% to 41.12\%. These results suggest that structured decomposition can improve consistency, while spatially grounded environmental risk recognition remains a challenging direction for building more reliable, evidence-based MLLMs in privacy- and security-sensitive authentication scenarios.

{
    \small
    \setlength{\baselineskip}{10.8pt}
    \setlength{\bibsep}{0pt}
    \bibliographystyle{ieeenat_fullname}
    \bibliography{main}
}

\end{document}